\documentclass[8.5pt,twoside,twocolumn]{article}
\usepackage[super,sort&compress,comma]{natbib} 
\usepackage[version=3]{mhchem}
\usepackage{balance}
\usepackage{times,mathptm}
\usepackage{sectsty}
\usepackage[dvips]{graphicx}  
\usepackage{lastpage}
\usepackage[format=plain,justification=justified,singlelinecheck=false,font=small,labelfont=bf,labelsep=space]{caption} 
\usepackage{fancyhdr}
\begin{document}

\thispagestyle{plain}
\fancypagestyle{plain}{
\fancyhead[L]{\includegraphics[height=8pt]{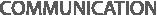}}
\fancyhead[C]{\hspace{-1cm}\includegraphics[height=20pt]{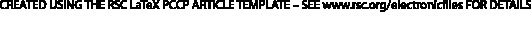}}
\fancyhead[R]{\hspace{10cm}\vspace{-0.25cm}\includegraphics[height=10pt]{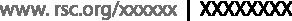}}
\renewcommand{\headrulewidth}{1pt}}
\renewcommand{\thefootnote}{\fnsymbol{footnote}}
\renewcommand\footnoterule{\vspace*{1pt}%
\hrule width 3.4in height 0.4pt \vspace*{5pt}} 
\setcounter{secnumdepth}{5}

\makeatletter 
\renewcommand\@biblabel[1]{#1}            
\renewcommand\@makefntext[1]%
{\noindent\makebox[0pt][r]{\@thefnmark\,}#1}
\makeatother 
\renewcommand{\figurename}{\small{Fig.}~}
\sectionfont{\large}
\subsectionfont{\normalsize} 

\fancyfoot{}
\fancyfoot[LO,RE]{\vspace{-7pt}\includegraphics[height=9pt]{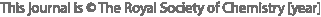}}
\fancyfoot[CO]{\vspace{-7.2pt}\hspace{12.2cm}\includegraphics{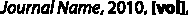}}
\fancyfoot[CE]{\vspace{-7.5pt}\hspace{-13.5cm}\includegraphics{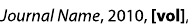}}
\fancyfoot[RO]{\footnotesize{\sffamily{1--\pageref{LastPage} ~\textbar  \hspace{2pt}\thepage}}}
\fancyfoot[LE]{\footnotesize{\sffamily{\thepage~\textbar\hspace{3.45cm} 1--\pageref{LastPage}}}}
\fancyhead{}
\renewcommand{\headrulewidth}{1pt} 
\renewcommand{\footrulewidth}{1pt}
\setlength{\arrayrulewidth}{1pt}
\setlength{\columnsep}{6.5mm}
\setlength\bibsep{1pt}

\twocolumn[
  \begin{@twocolumnfalse}
\noindent\LARGE{\textbf{Polymer-Enforced Crystallization of a Eutectic Binary Hard Sphere Mixture}}
\vspace{0.6cm}

\noindent\large{\textbf{Anna Kozina\textit{$^{a,\ddag}$}, Dominik Sagawe ($\dag$)\textit{$^{a}$}, Pedro D\'{\i}az-Leyva\textit{$^{a,b,\S}$}, Eckhard Bartsch\textit{$^{a,b,\ast}$} and Thomas Palberg\textit{$^{c}$}}}\vspace{0.5cm}

\noindent\textit{\small{\textbf{Received Xth XXXXXXXXXX 20XX, Accepted Xth XXXXXXXXX 20XX\newline
First published on the web Xth XXXXXXXXXX 200X}}}

\noindent \textbf{\small{DOI: 10.1039/b000000x}}
 \end{@twocolumnfalse} \vspace{0.6cm}

  ]

\noindent\textbf{We prepared a buoyancy matched binary mixture of polydisperse polystyrene microgel spheres of size ratio $\Gamma = 0.785$ at a volume fraction of $\Phi = 0.567$ just below the kinetic glass transition. In line with theoretical expectations, a eutectic phase behaviour was observed, but only a minor fraction of the samples crystallized at all. By adding a short non-adsorbing polymer we enforce \textit{inter}-species fractionation into coexisting pure component crystals, which in turn also shows signs of \textit{intra}-species  fractionation. We show that in formerly inaccessible regions of the phase diagram binary hard sphere physics are made observable using attractive hard spheres.}
\section*{}
\vspace{-1cm}




\footnotetext{\textit{$^{a}$~Institut f\"ur Makromolekulare Chemie, Albert-Ludwigs-Universit\"at Freiburg, D-79104 Freiburg, Germany}}
\footnotetext{\textit{$^{b}$~Institut f\"ur Physikalische Chemie, Albert-Ludwigs-Universit\"at Freiburg, D-79104 Freiburg, Germany}}
\footnotetext{\textit{$^{c}$~Institut f\"ur Physik, Johannes Gutenberg Universit\"at Mainz, D-55128 Mainz, Germany}}


\footnotetext{\ddag~Current address: Instituto Nacional de Investigaciones Nucleares, 52750 La Marquesa Edo.Mex., Mexico}
\footnotetext{\S~Current address: Departamento de F\'{\i}sica, Universidad Aut\'onoma Metropolitana, Iztapalapa, 09340 M\'exico, D.F., Mexico}

Binary colloidal mixtures are valuable model systems for fundamental studies of crystallization \cite{Bartlett1994Springer, Gasser2009JPCM, Lorenz2009JPCM}. Detailed predictions of their phase behaviour exist for hard sphere (HS) systems of different size ratio $\Gamma = R_{S} / R_{L}$ (where  $R_{i}$ are the radii of small ($S$) and large ($L$) spheres, respectively). These show a sequence of spindle to azeotropic to eutectic phase diagram types with decreasing $\Gamma$, vanishing miscibility in the crystal phase for $\Gamma < 0.85$ and a huge variety of crystal structures for compounds \cite{Bartlett1990JPCM, Kranendonk1991MolPhys, Cottin1993JCP, Eldridge1995MolPhys, Punnathanam2006JCP, Filion2009PRE2009}. For zero miscibility eutectics the \emph{inter}-species fractionation is expected to cause a pronounced slowing of nucleation \cite{Punnathanam2006JCP}. Polydispersities above $6\%$ destabilize the crystal phases \cite{Bolhuis1994PRE, Kofke1999PRE} (10$\%$ in 2D \cite{Pronk2004PRE}), hence additional \emph{intra}-species fractionation is expected for strongly polydisperse systems \cite{Xu2003JPC, Fasolo2003PRL}. Indications of polydispersity altered solidification dynamics have been observed in several HS experiments \cite{Schope2006PRE, Iacopini2009JCP} including subtle influences of the skewness of the size distribution \cite{Martin2003PRE} but direct evidence of \emph{intra}-species fractionation is still missing. While many predicted crystal structures have by now been observed \cite{Bartlett1994Springer, Gasser2009JPCM, BartletOttewill1990JCP, Eldridge1995MolPhys, Hunt2000PRE} and also the general sequence of phase diagram types was confirmed \cite{Meller1992PRL}, most parts of the phase diagram escape from a detailed experimental investigation; e.g., the exact locations of phase boundaries in compound forming systems have not yet been determined. Moreover, crystallization kinetics have not been obtained and, except for a recent 2D-study \cite{Geerts2010JCPM}, fractionation into coexisting $L$- and $S$-crystals has not yet been observed. One main reason for this is the interference of crystallization with the glass transition (GT) at large volume fractions \cite{Voigtmann2003PRE}. Furthermore, gravity seemingly enhances the trend to dynamically arrest the systems \cite{Zhu1997Nature}. In addition, it may lead to differential sedimentation and inhomogeneities in composition \cite{Lorenz2009JCP, Leocmach2010EPL}.

In the present paper we avoid sedimentation using a binary mixture of buoyancy matched microgel particles. We further exploit earlier observations on a variety of systems, in which both vitrification can be suppressed \cite{Pham2002Science, Chen2003PRE, Eckert2002PRL} and crystallization be accelerated \cite{Smits1990PT} by adding a short chained, non adsorbing polymer. Doing so, one moves from HS to attractive HS (AHS), which in principle may significantly alter both phase behaviour and crystallization kinetics \cite{Gast1984JCIS, Lekkerkerker1992PRL, Poon2000JPCM, Poon2002JPCM, Romero2004SM, Anderson2005Nature, Buzzacarro2007PRL, Zykova2010JCP}. However, using a binary AHS mixture at a size ratio of $\Gamma = 0.785$ we here give the first demonstration of the simultaneous precipitation of $S$- and $L$-crystals over the full range of compositions. Thus, we recover the zero miscibility eutectic solidification process expected for pure HS. Moreover, we provide experimental support for the theoretically expected additional \emph{intra}-species fractionation \cite{Fasolo2005JCP} and give a preliminary account of the crystallization kinetics. Further analysis suggests that the final degrees of crystallization as well as the time dependent conversion rates are controlled by both the polydispersity and the polymer-supported ability to form suitable composition fluctuations.

\begin{figure} [ht!]
\centering
\includegraphics[width=8.5cm]{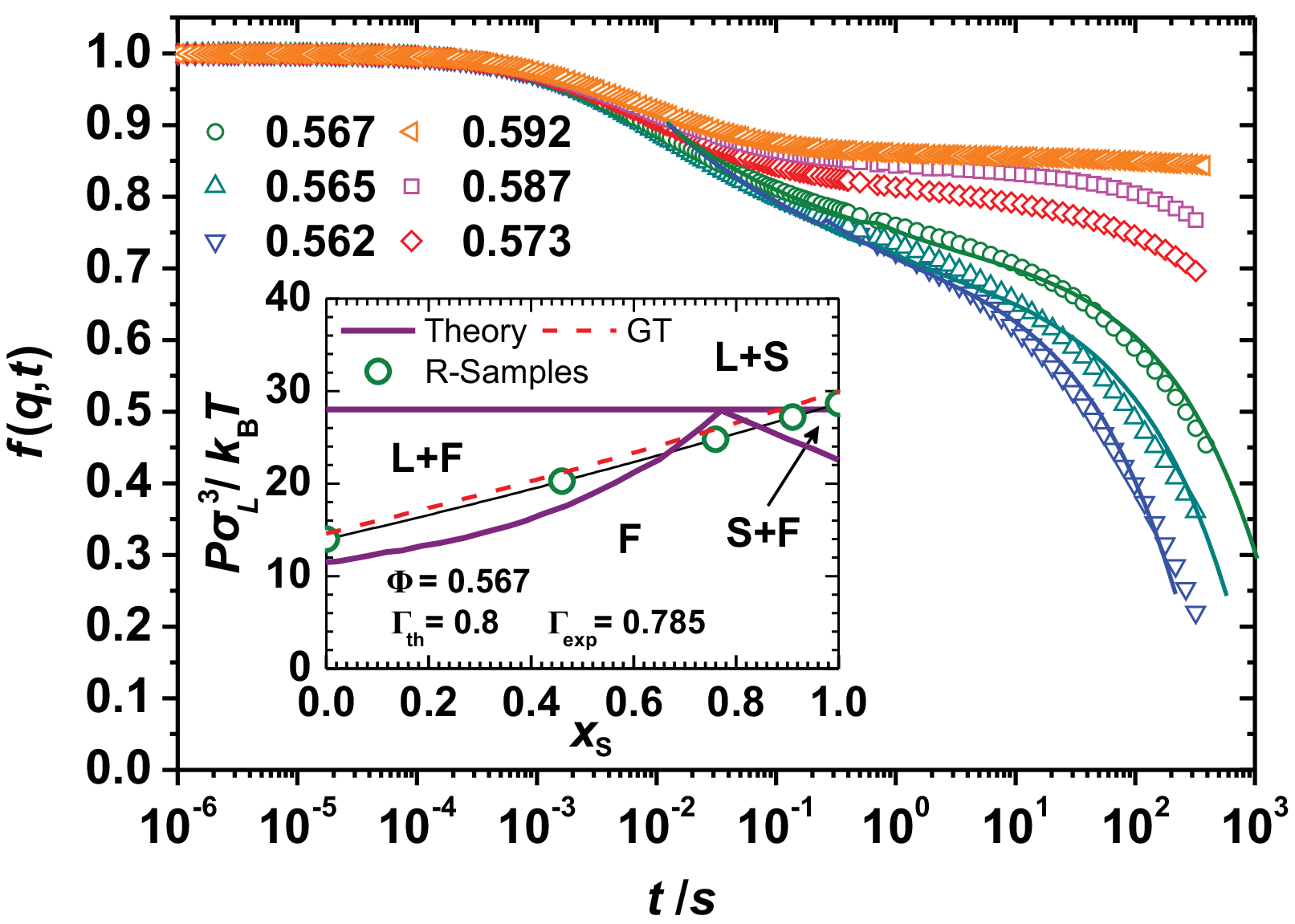}
\caption{(colour online) Main: intermediate scattering functions $f(q,t)$ $vs.$ $\log(t)$ for repulsive samples of eutectic composition $(x_S = 0.77)$ with increasing $\Phi$ from bottom to top. Inset: binary HS phase diagram in the pressure-composition plane. Thick solid lines: phase boundaries adapted from \cite{Punnathanam2006JCP} for $\Gamma_{th} = 0.8$. Thin solid line:  osmotic pressure for HS mixtures of experimental size ratio $\Gamma_h = 0.785$ calculated using the fundamental measure theory equation of state of \cite{Roth2010JPCM}. Dashed line: osmotic pressures along the GT-line ($\Phi = 0.573$). Circles: values for $R1$ - $R5$ ($\Phi = 0.567$) with partial volume fractions $\phi_S = \Phi_S / \Phi$ of 0, 0.26, 0.57, 0.83 and 1 corresponding to molar fractions $x_S = \rho_S / (\rho_S + \rho_L)$ of 0, 0.47, 0.77, 0.92 and 1 (where $\rho_i = \Phi_i / (4 \pi / 3) R_{h,i}^3$ is the particle number density).}
\label{fig1}
\end{figure}

Cross-linked polystyrene ($PS$) microgel spheres were synthesized by emulsion polymerization, cleaned, dried and re-suspended in 2-Ethylnaphtalene ($2EN$) as described previously \cite{Bartsch2002JNCS}. In this good solvent the particles swell to hydrodynamic radii of $R_{h,S} = (152 \pm 4)$ nm and $R_{h,L} = (193 \pm 3)$ nm giving a size ratio of  $\Gamma_{h} = 0.785$. The hardness of interaction was probed in oscillatory rheological measurements following the procedure described in \cite{Senff1999JCP}. Assuming an inverse power potential $U(r) \propto 1  /r^n$ we find $n = 40 \pm 2$ \cite{Kozina2009thesis} which is sufficiently steep to regard the particles as hard spheres. For both pure species the polydispersity index $\sigma = \sqrt{\left\langle R^2\right\rangle-\left\langle R\right\rangle^2} / \left\langle R\right\rangle$ is between 0.06 (TEM on dried particles) and 0.08 (form factor measurements). Sedimentation experiments following \cite{Paulin1990PRL} revealed their expected HS-like phase behaviour. The freezing concentration was then identified with the ideal HS value and the melting points scaled accordingly to obtain $\Phi_{M,S} = 0.528$ and $\Phi_{M,L} = 0.531$. Further, from the position of the first Bragg reflection of samples at coexistence the lattice spacing and from that effective radii were obtained as $R_{eff,L} = (170.5 \pm 2)$ nm and $R_{eff,S} = (126.2 \pm 3)$ nm (effective size ratio $\Gamma_{eff} = 0.74$) \cite{remark}. The inset of Fig. \ref{fig1} shows the expected phase diagram of monodisperse repulsive mixtures. Choosing the eutectic composition and increasing $\Phi$ we monitor the intermediate scattering functions, $f(q,t)$ (Fig. \ref{fig1} main). A Mode coupling GT is observed for $\Phi_{GT} = 0.573 \pm 0.002$ \cite{remarkGT}. For $\Gamma \leq 0.8$ the value of $\Phi_{GT}$ hardly varies with composition \cite{Voigtmann2003PRE}. The expected GT-line is shown dashed in the inset. Note that neither the eutectic point nor ($L$+$S$)-coexistence region are accessible.

\begin{figure*} [ht!]
\centering
\includegraphics[width=17.8cm]{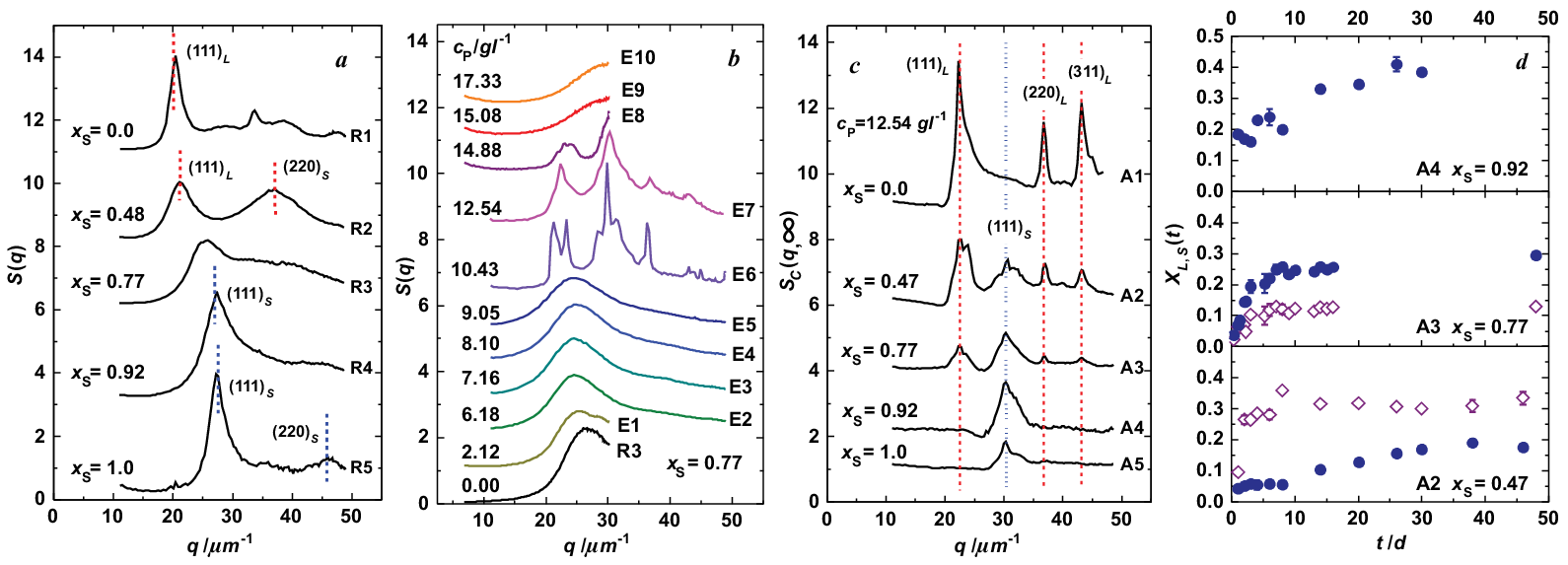}
\caption{(colour online) Scattering results. Curves are shifted for clarity. ($a$) Final structure factors of $R$-samples at different compositions as indicated ($\Phi = 0.567$). Dashed lines indicate location of Miller-indexed reflections averaged over several individual crystallites. ($b$) Final structure factors for the $E$-samples of $x_S = 0.77 = x_E$ and $\Phi = 0.567$ with indicated amounts of added polymer. ($c$) Crystal structure factors of $A$-samples with location of Miller-indexed reflections of the pure samples averaged over several individual crystallites (dashed lines) ($d$) Temporal development of the fraction of crystals for the three mixtures $A4$ - $A2$. Filled symbols denote $S$-crystals, open symbols denote $L$-crystals.}
\label{fig2}
\end{figure*}

Three Sample series $(R, E, A)$ were prepared to have $\Phi = 0.567 \pm 0.006$, just below $\Phi_{GT}$. $R$-samples to check the repulsive phase behaviour contain no polymer but vary in molar fraction of $S$-particles. $E$-samples of eutectic composition ($x_S = 0.77$) to monitor the effects of attraction contain increasing concentrations, $c_P$, of linear $PS$ ($R_g = 13.1$ nm, $M_W = 133$ kg/mol; PSS GmbH, Germany). Strongly attractive $A$-samples of different composition to explore ($L$+$S$)-coexistence have $c_P = 12.54$ g/l. Respecting the volume occupied by the colloids, the polymer concentration in the free volume is $c_{P,\mbox{free}} = 43$ g/l \cite{Illet1995PRE}, roughly 1.8 times larger than the overlap concentration $c_{P}^* = 3 M_W / (4 \pi N_A R_g^3) = 23.5$ g/l. (For more details on sample compositions see Fig. \ref{fig2})

Start of crystallization measurements ($t$ = 0) was defined by taking a sample off the homogenizing tumbler and mounting it on a static light scattering goniometer (modified Sofica, SLS Systemtechnik, Germany, $\lambda = 405$ nm, 1 mW laser illumination, angular range $25^{\circ}$ - $135^{\circ}$, resolution $1^{\circ}$). Structure factors, $S(q,t)$, were obtained by time resolved measurements of the scattered intensity, $I(q,t)$, dividing by the appropriate form factors, $P(q)$, obtained from a dilute sample and normalizing by the number density ratio: $S(q,t) = I _{conc}(q,t) \rho_{dilute} / P(q) \rho_{conc}$. Here $q = (4 \pi n_D /\lambda) \sin(\Theta/2)$ is the scattering vector with $n_D$ denoting the suspension refractive index. Zero time structure factors, $S(q,0)$, only show the meta-stable melt, while at later times $S(q,t)$ is a super-position of melt and crystal contributions. After two months most $S(q,t)$ stopped evolving. The crystal scattering contribution, $S_C(q,t)$ was obtained following \cite{Harland1997PRE}: $S_C(q,t) = S(q,t) - \beta S(q,0)$. The scaling factor  $\beta$ represents the fraction of remaining melt and the total fraction of crystals $X_C(t) = 1 - \beta(t)$. In this approximation we neglect the small changes of melt and crystal composition and thus scattering power with time.

The final $S(q)$ for the $R$-samples in Fig. \ref{fig2}($a$) are dominated by melt scattering. $R3$ did not crystallize within two years although $\alpha$-relaxation occurred within $10^3$ s. All other samples showed isolated small crystallites. The final crystallinities $X_C(\infty) \leq 0.2$ appear to be rather small. Presumably, values given by the lever rule applying to the coexistence regions are further diminished due to polydispersity \cite{Xu2003JPC}. In each sample we could isolate individual narrow Bragg-peaks appearing at different $q$. Their average scattering vectors (dotted lines) reasonably well coincide with the melt maxima. Since the calculated osmotic pressures of the mixtures \cite{Roth2010JPCM} are below the eutectic pressure $P_E = 28 k_B T(2R_L)^{-3}$ at $\Gamma = 0.8$ \cite{Punnathanam2006JCP}, our findings are qualitatively consistent with the expected eutectic phase behaviour and with previous results on a $3\%$ polydisperse mixture with $\Gamma = 0.72$ \cite{Hunt2000PRE}.

The $E$-samples in Fig. \ref{fig2}($b$) show fluid order up to $c_P = 9.05$ g/l and a minimum of the structural relaxation time at $c_P = 8.10$ g/l. For $c_P \geq 10.43$ g/l we observe crystallization, while for $c_P \geq 15.08$ g/l the samples initially are amorphous again. Their $f(q,t)$ show plateau-like features. $E9$ and $E10$ finally started crystallizing after two years. We then varied the composition with $c_P$ above the crystallization threshold. Fig. \ref{fig2}($c$) shows the $S_C(q,\infty)$ of the $A$-samples with $c_P = 12.54$ g/l. Large numbers of small crystallites appeared within days and grew to final sizes (0.4 - 0.8 mm) within several weeks. Absence of additional peaks and shifts in peak positions with $x_s$ excludes the formation of compounds and of substitutional crystals. By comparison to the $S_C(q,\infty)$ of the two pure samples we identify $A2$ and $A3$ to show crystals of both species, while $A4$ contains $S$-crystals only. This simultaneous occurrence of both crystal species shows that the systems are above the eutectic pressure (c.f. Fig. \ref{fig1}). Within experimental uncertainty the observed lattice constants and thus the particle number densities $n$ equal those of the pure samples. Using the latter and the effective radii, $R_{eff,i}$, we obtain the volume fractions for both coexisting crystal species to be $\Phi = 0.73 \pm 0.02$ for all samples. This is remarkably close to the maximum packing fraction obtainable in a close packed structure. At this size ratio similar compressions  have so far only been reached, when additional gravitational compression of sedimented crystals occurred \cite{Buzzacarro2007PRL}.

From time resolved measurements the fraction of crystalline material, $X(t)$, was obtained by integrating the intensity in the corresponding peak regions: $X_{L,S}(t) = c\int dq S_C(q)$, choosing $c$ such that $X_L(t) + X_S(t) = X_C(t) = 1 - \beta(t)$. Fig. \ref{fig2}($d$) shows that at $x_S = 0.92$ about $20\%$ of the $S$-particles form $S$-crystals already at early times which grow for about a month; at $x_S = 0.77$, $10\%$ of each species form crystals simultaneously; the final crystallinity is reached after 10 days; for $x_S = 0.47$ starting again from about $10\%$ of each species, $S$-crystal conversion accelerates after $L$-crystals have nearly reached their final crystallinity. In all samples the degree of crystallization $X_C(\infty)\approx$ 0.4 - 0.6 is considerably enhanced over that reached in the $R$-samples. Moreover, at this $c_P$ the final $X_S$ of 0.42, 0.4, 0.3 and 0.2 roughly scale with the number fractions of 1.0, 0.92, 0.77 and 0.47, showing that about $40\%$ of the $S$-particles are able to solidify via \textit{intra}-species fractionation. Similarly, for the $L$-particles we find $X_L$ to be 0.13, 0.34 and 0.6 for $(1 - x_S)$ being 0.23, 0.53 and 1.0, i.e. more than half of the particles crystallize.

Summarizing, we find the theoretically expected eutectic phase behaviour for a purely repulsive HS mixture of $\Gamma = 0.8$ and $\Phi = 0.567$ i.e. fluid-solid coexistence or pure fluid \cite{Bartlett1990JPCM, Kranendonk1991MolPhys, Cottin1993JCP, Eldridge1995MolPhys, Punnathanam2006JCP}, but access to both the eutectic point and the ($L$+$S$)-coexistence region is blocked by a kinetic glass transition. Upon addition of non adsorbing polymer, however, the eutectic point in terms of polymer concentration can be located at $9.05$ g/l $< c_{P,E} \leq 10.43$ g/l. As the pressure, $P$, surmounts the eutectic pressure $P_E$, i.e. the freezing pressure, $P_F(x_E)$, measured at the eutectic composition, ($L$+$S$)-coexistence is observed for all compositions. Recent work on one component AHS \cite{Buzzacarro2007PRL, Zykova2010JCP} showed that both $P(\Phi)$ and $P_F$ shift to lower values upon increasing the attraction. Our work suggests that for binary AHS $P_F$ drops faster than $P(\Phi = 0.567)$. However, theoretical predictions for composition dependent equations of state and the phase diagram of binary AHS are eagerly awaited to quantitatively rationalize our observations in a plot analogous to Fig. \ref{fig1}.

Being above $P_E$, our mixtures should convert completely to $S$- or $L$-crystals. Instead, we find a pronounced amorphous background signal and broad pyramid-shaped peaks composed from a superposition of a large number of individual Bragg reflections located at slightly different $q$. In addition to this overall incomplete conversion we observe an absence of $L$-crystals at large $x_s$ (already seen in \cite{Hunt2000PRE, Leocmach2010EPL}), an increased initial formation of $S$-crystals in $A4$, their delayed formation in $A2$. This supports theoretical expectations that conversion proceeds via combined \emph{inter}- \cite{Punnathanam2006JCP} and \emph{intra}-species fractionation \cite{Fasolo2005JCP}. In both cases the conversion efficiency should be correlated with the possibility to form combined density and composition fluctuations. The addition of polymer generally enhances the dynamics of such fluctuations \cite{Eckert2002PRL} as long as the system remains un-trapped by the attractive glass. However, conversion efficiency also relies on the availability of similar sized partner particles to form a stable fluctuation. Their concentration in the remaining melt changes in time. Majority species and component particles crystalize first. The enrichment of the minority particles changes the corresponding fluctuation spectra and allows further conversion. However, particles from the wings of a size distribution or of a very diluted species simply stay too dispersed to crystallize.

In this study we have recovered some interesting aspects of HS-crystallization in an AHS system. In future, the controlled use of depletion attraction in formerly inaccessible regions of the phase diagram may allow detailed and quantitative studies of the phase behaviour at other size ratios, kinetics of compound formation or other interesting problems of condensed matter physics.

We thank H. Moschallski for particle synthesis and purification. Financial support of the DFG (Pa459/8, 12-14, 16; Ba1619/2; SFB 428; SFB TR6), the EU (CT-2003-504 712), and the Graduate School of Excellence, Material Science in Mainz, is gratefully acknowledged.

\balance

\footnotesize{

 }

\end{document}